\documentclass[twocolumn,pra,aps,]{revtex4-1}

\usepackage{amssymb,amsmath,amsfonts,amstext,graphics,graphicx, subfigure}

\usepackage{color}
\newcommand{\bs}[1]{\ensuremath{\boldsymbol{#1}}} 
\newcommand{\p}{\partial}

\def\cale{\mathcal{E}}

\def\calu{\mathcal{U}}

\def\E{{\rm I\kern-.1567em E}}
\def\A{{\rm I\kern-.1567em A}}
\def\P{{\rm I\kern-.1567em P}}
\def\V{{\rm I\kern-.1567em V}}
\def\bq{\begin{equation}}
\def\eq{\end{equation}}
\def\bqy{\begin{eqnarray}}
\def\eqy{\end{eqnarray}}

\def\al{\alpha}

\def\De{\Delta}

\def\ga{\gamma}

\def\si{\sigma}

\def\bfj{\mathbf{J}}

\def\bfV{\mathbf{V}}

\def\bfv{\mathbf{v}}

\begin{document}

\title{On  energy conservation in extended  magnetohydrodynamics}
\author{Keiji Kimura$^1$ and P. J. Morrison$^2$}
\affiliation{$^1$ Research Institute for Mathematical Sciences,Kyoto University, Kitashirakawa Oiwake-cho, Sakyo-ku, Kyoto 606-8502, Japan. \\ $^2$Department of Physics and Institute for Fusion Studies, University of Texas at
Austin, Austin, TX 78712-1060}
\date{\today}

\begin{abstract}
A systematic study of  energy conservation for  extended magnetohydrodynamic (MHD)  models that include Hall terms and electron inertia is performed.  It is observed  that commonly used models do not conserve energy in the ideal limit, i.e., when viscosity and resistivity are neglected.   In particular, a term  in the momentum equation that  is often  neglected  is seen to be  needed for  conservation of  energy.   

\bigskip 


\bigskip


\end{abstract}

\maketitle

%%%%%%%%%%%%%%%%%%%%%%%%
%%%%%%%%%%%%%%%%%%%%%%%%
\section{Introduction}
\label{Introduction}

Ideal  magnetohydrodynamic (MHD) has a long history of wide application  to  types of plasmas,  including those of relevance in astrophysics, geophysics, and nuclear fusion science.  However, it is well-known to plasma physicists that ideal MHD is deficient in many respects and that some of these deficiencies  are  accounted for by extending Ohm's law as in the early work of Refs.~\cite{1959-Luest,spitzer} (see also \cite{50schluet,51schluet}).  And, it is also well-known that the inclusion of additional terms in Ohm's law breaks the frozen flux condition of ideal MHD and gives rise to  specific regions where magnetic reconnection takes place and that this phenomenon is important for energy transfer. Reconnection is most well-studied when it is induced by  resistivity  (e.g.\ \cite{1975-Vasyliunas}),  but there is significant work on the Hamiltonian reconnection afforded by other effects such as electron inertia (e.g.\  \cite{1993otavani,pegoraro00,hirota}).  Because a variety of effects can be important,  many researchers have investigated various extended MHD models  both analytically and numerically, and many  reduced models based on geometric reduction have been used  as well (e.g.,  \cite{1994schep, 1998pegoraro,tassi08}).   

Different versions  of extended MHD models have been implemented, e.g.,  \cite{spitzer,1975-Vasyliunas,1999-Bhattacharjee_etal,2008-Fitzpatrick,1982-Freidberg},  and  some of the relationships between these models and their limitations  seem  to be unknown.  All of these models differ  only  in the choice of the generalization of  Ohm's law,  with the momentum equation being  the same  as that for usual MHD. We will see that in some instances, energy conservation requires modification of the momentum equation. 

The  goal of this paper is  to sort out which extended MHD models conserve energy and which do not.   Upon returning to the original two-fluid derivation of extended MHD of L\"{u}st \cite{1959-Luest} we find  that energy conservation requires a term that is often  neglected in the momentum equation, and that retention of this term is consistent with the appropriate ordering.  Various reductions of the full extended MHD model are investigated,  including Hall and Inertial MHD.

This paper is organized as follows.  In Sec.~\ref{sec:models-cons} we describe briefly extended MHD and generalize the thermodynamics of the fluid to allow  for  more general equations of state with  the inclusion of  electron pressure and anisotropic pressure of the form of Chew, Goldberger, and Low \cite{cgl}.    Next, in Sec.~\ref{sec:Econs} we begin our discussion of energy conservation.  We first consider Hall MHD (HMHD), which is a well-known consistent model in its own right, and verify  its energy conservation including the generalized thermodynamics.   Then,  we introduce inertial MHD (IMHD)  by employing an ordering of the full extended MHD  in which the terms of Ohm's law of HMHD are  dominated.  Thus,  we are able to treat the energy conservation of IMHD independently,  but our results apply to the full extended MHD model without the ordering.   Since various IMHD-like models in the literature do not conserve energy, these are of main concern.   For this reason, in Sec.~\ref{sec:class-cons},  we  systematically determine which MHD  models with electron inertia conserve energy and which do not -- the results are summarized in Table \ref{table:classification_IMHD}. 
Finally, we conclude  in Sec.~\ref{sec:conclu}, where we discuss some limitations and possibilities.

%%%%%%%%%%%%%%%%%%%%%%%%
%%%%%%%%%%%%%%%%%%%%%%%%
\section{Extended MHD  and thermodynamics}
\label{sec:models-cons}

In this section we first state the extended MHD model.   As is well-known, such a one-fluid model can be derived from  kinetic theory (see e.g.\  \cite{spitzer,2008-Fitzpatrick,1984-Chen}), but we begin with the results  of L\"{u}st \cite{1959-Luest},   who appears to be the first to derive the generalized Ohm's law for a one-fluid model by  adding and subtracting individual electron and ion fluid equations, enforcing quasineutraility, and expanding in the smallness of the electron mass.  His derivation yields a term in the one-fluid momentum equation that is often neglected and is necessary for energy conservation.  Next, we  extend  L\"{u}st's model by   completing the thermodynamics and, in addition, we show how one can incorporate  anisotropic pressure into the thermodynamics.

%%%%%%%%%%%%%%%%%%%%%%%%
\subsection{Extended MHD}

The assumptions of quasineutrality and smallness of  the electron mass compared to the ion mass leads to  a model that we will refer to as {\it extended MHD}.  It is given by the following:
\medskip

\noindent the continuity equation,
\bq
\frac{\p \rho}{\p t} =  -  \nabla\cdot(\rho \bs{V})\,,
\label{1cont}
\eq
the momentum equation, 
\begin{align}
\rho \left( \frac{\p \bs{V}}{\p t} + (\bs{V}\cdot\nabla)\bs{V}\right)
 &= - \nabla p + \bs{J}\times\bs{B}
 \label{1mom}\\
  &\hspace{.5 cm}  - \frac{m_e}{e} (\bs{J}\cdot\nabla)\frac{\bs{J}}{en}\,,
\nonumber
\end{align}
and the generalized Ohm's law
\begin{align}
\bs{E} + \bs{V}\times\bs{B}- \frac{\bs{J}}{\sigma}
 &= 
   \frac{1}{en}
    (\bs{J}\times\bs{B} - \nabla p_e) \nonumber\\
 &\hspace{.20 cm}
   + \frac{m_e}{e^2 n}
     \left[ \frac{\p \bs{J}}{\p t}
            + \nabla\cdot( \bs{V}\bs{J} + \bs{J}\bs{V}) \right] \nonumber\\
 &
 \hspace{1.25 cm} - \frac{m_e}{e^2 n} (\bs{J}\cdot\nabla) \frac{\bs{J}}{en}\,,
  \label{1ohm}
\end{align}
where $\rho$ is the mass density of plasma,
$\bs{V}$ the bulk velocity,
$p$ the pressure,
$\bs{B}$ the magnetic field,
$\bs{J}=\nabla\times \bs{B}/\mu_0$ the current density,
$m_e$ the electron mass,
$e$ the elementary charge,
$n$ the number density of each species of charged particles,
$\bs{E}$ the electric field,
$\sigma$ the conductivity, and 
$p_e$ the electron pressure.

Although this generalized Ohm's law arises upon  subtracting the individual  electron and ion fluid momentum equations, with velocity fields  $\mathbf{v}_e$ and $\mathbf{v}_i$, respectively, and contains  electron momentum dynamics,  we will refer to this system of Eqs.~\eqref{1cont},  \eqref{1mom}, and \eqref{1ohm}, supplemented by the thermodynamics of Sec.~\ref{ssec:thermo},  as a single fluid model.

%%%%%%%%%%%%%%%%%%%%%%%%
\subsection{Extended thermodynamics}
\label{ssec:thermo}

In the original article  of L\"{u}st \cite{1959-Luest}, and to our knowledge elsewhere, the thermodynamics of the two-fluid to one-fluid extended MHD reduction has not been treated in full generality.   L\"{u}st assumes polytropic laws for $p_i$ and $p_e$, and constructs the extended MHD pressure as $p=p_i+p_e$, in accordance with   Dalton's law on the addition of partial pressures.   Because of quasineutrality, $\rho=\rho_i + \rho_e= m_i n + m_e n= mn$, where $m=m_i+m_e$.   

Here we generalize this by using entropy as  the second thermodynamic variable for each species.   With this choice, the thermodynamics of species $\al\in\{e,i\}$  is determined from an internal energy function $\calu_{\al}(\rho_{\al}, s_{\al})$,  the internal energy per unit mass $m_{\al}$,  where  $s_{\al}$ is the entropy per unit  mass $m_{\al}$.  Since $1/\rho_{\al}$ is the specific volume, this thermodynamic representation is the one in terms extensive variables, with the intensive quantities   determined by 
\bq
T_{\al}=\frac{\p \calu_{\al}}{\p s_{\al}} \quad\mathrm{and}\quad p_{\al}=\rho_{\al}^2 \frac{\p \calu_{\al}}{\p \rho_{\al}}\,.
\eq
For isothermal processes $\mathcal{U}_{\al}=\kappa(s_{\al})\ln(\rho_{\al})$, while for 
a polytropic equation of state,  $\calu_{\al}=\kappa(s_{\al}) \rho_{\al}^{\ga-1}/(\ga-1)$, whence  $p_{\al}=\kappa(s_{\al})\rho_{\al}^{\ga}$.  For these choices,   one can substitute the variable $p_{\al}$ in lieu of $s_{\al}$,  as is more common in plasma physics.

For the ideal, energy conserving, fluid the Fourier and other heat flux terms are  dropped and the two entropies obey the advection equations, 
\bq
\frac{\p s_{\al}}{\p t} +\mathbf{v}_{\al} \cdot \nabla s_{\al}=0\,.
 \eq
Since entropy is extensive it is natural to introduce the total entropy for  one-fluid extended MHD as follows:
\bq
s=(m_i s_i + m_e s_e)/m\,.
\label{ssum}
\eq 
In addition,  if  Hall MHD is to have a complete set of thermodynamic variables, one must retain the electron entropy, $s_e$.   Using $\bs{V}=(m_i \bs{v}_i + m_e \bs{v}_e)/ m $ and $\bs{J}= en(\bs{v}_i-\bs{v}_e)$, a simple calculation gives
\bq
\frac{\p s}{\p t} =  -  \bs{V}\cdot\nabla  s\,,
\label{ent}
\eq
and
\bq
\frac{\p s_e}{\p t} =  -  \bs{V}\cdot\nabla  s_e  +   \frac1{en} \bs{J}\cdot\nabla s_e \,,
\label{els}
\eq
where we have dropped terms of order $m_e/m$.  

Again, appealing to the extensive property of energy, the  total internal energy per unit volume of the two species is given by 
\bq
\rho \calu= \rho_i\calu_i(\rho_i,s_i) + \rho_e\calu_e(\rho_e,s_e) \,.
\label{inten}
\eq
Because of quasineutrality, $n$ is the only density variable; for our purposes it is sufficient to rewrite this, again correct to order $m_e/m$, as 
 \bq
 n\mathfrak{U}(n,s,s_e)= n\mathfrak{U}_i(n,s)
  + n\mathfrak{U}_e(n,s_e)\,,
 \label{ppU}
 \eq
where $m\calu=:\mathfrak{U}$ and $m_{\al}\calu_{\al}=:n\mathfrak{U}_{\al}$.  Evidently, the pressures satisfy  $p=\rho^2 \p \calu/\p \rho=n^2 \p \mathfrak{U}/\p n= p_i + p_e$.   Consistent with the derivation of extended MHD, we can expand \eqref{ppU} in the smallness of $m_e/m_i$.   Because the only thermodynamic deviation from single fluid MHD occurs in the Hall term $\nabla p_e$, it is sufficient to retain only the leading order in this expansion in order to ensure energy conservation.  This will be shown in Sec.~\ref{ssec:EconsH} where we examine the total energy conservation for Hall MHD. 

Since it is not widely known, we review here the thermodynamics for anisotropic pressure in a magnetofluid model, which to our knowledge was first given in \cite{aip82} (see also \cite{HMM}).   The generalization follows upon adding $B=|\bs{B}|$ as an additional thermodynamic variable, i.e., now $\calu(\rho, s, B)$, and the parallel and perpendicular pressures are given by
\bq
p_{||}=\rho^2 \frac{\p \calu}{\p \rho}
\quad \mathrm{and}\quad
\De p=- \rho B \frac{\p \calu}{\p B}\,.
\label{thermoB}
\eq
where $\De p= p_{||} - p_{\perp}$.

Expressions \eqref{thermoB} can be seen to be consistent with  the natural  extensive thermodynamic variables  $(\rho^{-1}, s, B)$, all being specific quantities.  The intensive thermodynamic dual variables associate with this set are
$(p_{\parallel}, T, M)$ with the magnetization $M$ being given by 
\bq 
M:=\frac{\De p}{\rho B}=-\frac{\p \calu}{\p B}\,.
\label{M}
\eq
In \eqref{M},  $\De p=p_{||}- p_{\perp}$  is related to the work done when magnetic flux is held fixed.  Note the magnetic moment $\mu=mv^2_{\perp}/(2B)$;  thus the magnetization per unit volume is
$\mathcal{M}= n mv^2_{\perp}/(2B)$, or one can argue macroscopically $\mathcal{M}\sim p/B$.  Therefore, the specific magnetization would be $M\sim p/(\rho B)$;  thus (\ref{M}) makes sense as a relative magnetization. 
 Alternatively, one can use $H$ as the variable thermodynamically conjugate to $B$ by introducing the `total' energy, 
$ \calu_{tot}=\rho \calu +  {B^2}/{2\mu_0}$.  Then,  the  conjugate to $B$ is given by 
${\p \calu_{tot}}/{\p B}= H= -\mathcal{M} + B/\mu_0$, 
as expected.  In \cite{aip82} it was shown that this way of introducing anisotropy is energy conserving.    For simplicity, in  the following we will restrict to isotropic pressure, but the generalization is straightforward.

%%%%%%%%%%%%%%%%%%%%%%%%
%%%%%%%%%%%%%%%%%%%%%%%%

\section{Energy conservation}
\label{sec:Econs}

Because the examination of general energy conservation of extended MHD is complicated, we divide the calculation into two parts.  We first consider HMHD and then its complement  IMHD.  Results for total energy follow upon superposing the calculations.  In addition, we give an ordering for IMHD, an energy conserving model in its own right. 
 
%%%%%%%%%%%%%%%%%%%%%%%%
\subsection{Hall MHD}
\label{ssec:EconsH}

Setting $m_e=0$ in  \eqref{1mom} and  \eqref{1ohm} gives resistive HMHD.  Since electron inertia is absent, the energy is expected to be composed of the sum of kinetic, internal, and magnetic, i.e.
\bq
H_{H} := \int_{D} \!d^3x\left( \rho
       \frac{  |\bs{V}|^2}{2}  + \rho\, \calu
       + \frac{|\bs{B}|^2}{2 \mu_0}
     \right) \,,
     \label{hallE}
\eq
where it remains to determine the function $\calu$ that will ensure conservation of $H_{H}$ when the resistivity $\si^{-1}$ is set to zero.  Determination of $\calu$ will be tantamount to the determination of   the entropy dynamics. 

Upon calculating $d H_{H}/dt$ it is readily seen that the  MHD terms cancel as usual and that  the Hall term $\bs{J}\times \bs{B}$ produces the energy flux $\bs{B}\times(\bs{B}\times\bs{J})/(\mu_0 e n)$.  Consequently, only the thermodynamic terms are of concern for the energy of \eqref{hallE} to satisfy a conservation law.  Assuming $\calu(\rho, s, s_e)$ and $p=\rho^2\p  \calu/\p \rho$, one obtains the usual internal energy flux of MHD, $(p + \rho \calu)\bs{V}$.  Thus, we are left with the following upon neglect of surface terms:
\bq
\frac{d H_{H} }{dt}= \int_{D} \!d^3x\Big(
  \frac1{en} \bs{J}\cdot \nabla p_e  +  \frac{\rho}{en} \frac{\p \calu}{\p s_e}\bs{J}\cdot \nabla s_e
  \Big)
\,,
\label{Hsm}
\eq
where use has been  made  of  \eqref{ent} and  \eqref{els}.

For barotropic electron pressure,   $\calu$ has no dependence on the electron entropy  $s_e$ and, consequently,  only   the first term of \eqref{Hsm}  is present and  $\nabla p_e/(en)=\nabla(\rho\calu)'/e$, where prime denotes $d/dn$.   Then, upon integration by parts we obtain
\bq
\frac{d H_{H} }{dt}= \int_{D} \!d^3x\, ( 
 \bs{J}\cdot \nabla (\rho\calu)'/e)= \int_{D} \!d^3x\, 
 \nabla\cdot(\bs{J} (\rho\calu)'/e)\,, 
 \label{dotHH}
 \eq
using $\nabla\cdot\bs{J}=0$.  Thus, for this case  the energy flux is $-\bs{J}(\rho\calu)'/e $.

The barotropic  model is incomplete since electron pressure can change at fixed electron density.  To account for this we  have included the  electron entropy in the dynamics via $\calu$.  Using \eqref{ppU} with $p_e=n^2\p \mathfrak{U}_e/\p n$ and $\rho\p \calu/\p s_e=  n\p \mathfrak{U}_e/\p s_e$, \eqref{Hsm} becomes
\bqy
\frac{d H_{H} }{dt}&=&\frac1{e} \int_{D} \!d^3x\bigg[
  \frac1{n} \bs{J}\cdot \nabla\left( n^2\frac{\p \mathfrak{U}_e}{\p n}\right) +   \frac{\p \mathfrak{U}_e}{\p s_e}\bs{J}\cdot \nabla s_e
  \bigg]
\,,
\nonumber
\\
&=&\frac1{e} \int_{D} \!d^3x\bigg[ \nabla\cdot \left(\bs{J}\, n\frac{\p \mathfrak{U}_e}{\p n}\right) 
   +   \frac{\p \mathfrak{U}_e}{\p n}\bs{J}\cdot \nabla n
   \nonumber\\
   &&\hspace{4 cm} +  \frac{\p \mathfrak{U}_e}{\p s_e}\bs{J}\cdot \nabla s_e\bigg]
      \nonumber\\
      &=&\frac1{e} \int_{D} \!d^3x  \nabla\cdot \left[\bs{J}\left( n\frac{\p \mathfrak{U}_e}{\p n}
      +  \mathfrak{U}_e\right) \right]\,,
\label{Hsm2}
\eqy
yielding $-\bs{J}\left( n{\p \mathfrak{U}_e}/{\p n} +  \mathfrak{U}_e\right)/e$ as another contribution to the energy flux. 

Thus,  we conclude that HMHD has the integrand of \eqref{hallE} as an energy density, say $\cale_H$, and this quantity  satisfies a conservation law of the form $\p \cale_H/\p t + \nabla\cdot \bs{J}_H=0$ for an energy flux $ \bs{J}_H$ given by
\bq
 \bs{J}_H=  \bs{J}_{MHD} + \frac{B^2}{\mu_0 e n} \bs{J}_{\perp}
 -\frac1{e}\left( n\frac{\p \mathfrak{U}_e}{\p n} +  \mathfrak{U}_e\right) \bs{J} 
 \,,
 \eq
where $\bs{J}_{MHD}$ is the usual MHD energy  flux. 

%%%%%%%%%%%%%%%%%%%%%%%%
\subsection{Inertial MHD}
\label{ssec:EconsI}

Now let us consider the remaining  terms of extended MHD.   The last term on the right-hand-side of the momentum equation \eqref{1mom} exists due to  electron inertia, and we will see that its retention is crucial when electron inertia terms are included in Ohm's law.    We will call  the second term on the left-hand-side
of the generalized Ohm's law \eqref{1ohm} the  ``nonlinear term,''
the third term on the left-hand-side the  ``collision term,''  the first line on the right-hand-side  the ``Hall term,''
and the remaining terms on the right-hand-side the  ``electron inertia terms.''
To compare the size of these terms,
we use the following dimensionless numbers: 
\begin{align*}
R_M &:= \frac{\mbox{Nonlinear term}}{\mbox{Collision term}}
      = \sigma \mu_0 U L, \\
C_H  &:= \frac{\mbox{Hall term}}{\mbox{Collision term}}
      = \frac{\sigma B}{e n}, \\
C_I  &:= \frac{\mbox{Electron inertia term}}{\mbox{Collision term}}
      = \frac{\sigma m_e}{e^2 n \tau},
\end{align*}
where $U$, $L$, $B$, and $\tau$ are the characteristic
velocity scale, length scale, magnitude of magnetic field, 
and time scale of current change, respectively.  Here  
$R_M$ is usual magnetic Reynolds number or Lundquist number as it is sometimes called. The two Hall terms are comparable in size if $B^2\sim p_e$, i.e., $\beta_e\sim 1$.  

For IMHD  we focus on the situation where  the electron inertia term is  larger than
the collision  and  Hall terms;  however, the nonlinear term is still considered to be  comparable with the electron inertia term, that is,
\begin{align*}
R_M \gg 1,~ C_I \gg 1,~ \frac{C_I}{C_H} \gg 1.
\end{align*}
Since the last inequality is equivalent to 
\begin{align*}
\frac{C_I}{C_H} = \frac{m_e}{eB} \frac{1}{\tau}
 = \frac{1}{\Omega_{e} \tau}\gg 1,
\end{align*}
where $\Omega_{e}$ is the electron gyro-frequency,
this relation can be interpreted as saying that 
 the characteristic time scale of the current change
is much shorter than the gyro-period of the electron.
%The magnetic reconnection is considered to occur very fast
%so we expect that the electron inertia term can be effective
%near the magnetic reconnection region.

With the above ordering, extended MHD reduces to the  IMHD model  given by  the following set of equations:
\begin{align}
 \frac{\p \rho}{\p t} &= - \nabla\cdot(\rho \bs{V})   ,
   \label{eq:continuity} \\
\rho \left( \frac{\p \bs{V}}{\p t} + (\bs{V}\cdot\nabla)\bs{V}\right)
 &= - \nabla p
   + \bs{J}\times\bs{B}
   \nonumber\\
   &\hspace{.5 in}
   - \epsilon~ \frac{m_e}{e} (\bs{J}\cdot\nabla)\frac{\bs{J}}{en},
   \label{eq:momentum} \\
\bs{E} + \bs{V}\times\bs{B}
 &= \epsilon~ \frac{m_e}{e^2 n}
      \left[ \frac{\p \bs{J}}{\p t}
            + \nabla\cdot( \bs{V}\bs{J} + \bs{J}\bs{V}) \right]
               \nonumber\\
   &\hspace{.5 in}
   - \delta \frac{m_e}{e^2 n} (\bs{J}\cdot\nabla) \frac{\bs{J}}{en},
   \label{eq:ohm} \\
\frac{\p s}{\p t} &=  -  \bs{V}\cdot\nabla  s,
   \label{eq:adiabatic}
\end{align}
where $s$ is the entropy per unit mass of the plasma
and the last equation means the plasma is adiabatic.   Note, we have artificially inserted book keeping parameters 
 $\epsilon$ and $\delta$ in order to identify terms --  in reality, both of these parameters have value unity. 
The above equations are to be solved with the pre-Maxwell's equations,
\begin{equation}
\nabla\times\bs{E} = - \frac{\p \bs{B}}{\p t} \qquad {\rm and }\qquad
\nabla\times\bs{B} = \mu_0 \bs{J}\,,
\end{equation}
with the initial condition $\nabla\cdot\bs{B} = 0$.  Note,  consistent with  quasineutrality and the neglect of the Maxwell displacement current,  the current density is solenoidal,  $\nabla \cdot \bs{J} = 0$.

We stress  that the energy conservation results we obtain do not depend on the IMHD ordering, but exist with the inclusion of the Hall terms, provided one extends the thermodynamics as in Sec.~\ref{ssec:EconsH}.  For IMHD one only need consider $\calu(\rho, s)$, but it could also be generalized to include $B$.

Upon considering  a candidate  energy of this IMHD model by taking the scalar product of $\bs{V}$ and the momentum equation,
the scalar product of $\bs{J}$ and the generalized Ohm's law,
and using  the pre-Maxwell equations, we obtain the following energy relation:
\begin{align}
&\frac{\p}{\p t}\left(
       \frac{\rho }{2}|\bs{V}|^2 + \rho\,  \mathcal{U}
       + \epsilon \frac{m_e}{e^2 n}\frac{|\bs{J}|^2}{2}
       + \frac{|\bs{B}|^2}{2 \mu_0}
     \right) \nonumber\\
 &~~~~
   + \nabla\cdot \left[
      \left( \frac{\rho}{2}|\bs{V}|^2 + p + \rho \,\mathcal{U}           
       + \epsilon \frac{m_e}{e^2 n} \frac{|\bs{J}|^2}{2} \right) \bs{V}
       \right.       \label{Hcon}\\
 &~~~
     \left.
      + \epsilon \frac{m_e}{e^2 n} (\bs{V}\cdot\bs{J})\bs{J}
      - \delta \frac{m_e}{2 e^3 n^2} |\bs{J}|^2 \bs{J}
      + \frac{\bs{E}\times\bs{B}}{\mu_0}
     \right] = 0.
\nonumber
\end{align}
Observe the new term of the energy density of (\ref{Hcon}),   $ {m_e}{|\bs{J}|^2}/({2}{e^2 n})$, which arises from electron inertia and  represents the electron kinetic energy density. 
Note that, from the generalized Ohm's law, because of the dependence of $\bs{E}\times \bs{B}$ in the energy flux of \eqref{Hcon}, the flux includes the time derivative term
$\epsilon {m_e}{\p \bs{J}}/{\p t}/({e^2 n})$, so the above formulation is not  in the usual conservation form.
However, upon  integrating the above energy relation over  the whole domain $D$ with appropriate boundary conditions,
it is revealed that the total energy $H$, which is defined as
\begin{align*}
H := \int_{D} \!d^3x\left(
       \frac{\rho}{2}  |\bs{V}|^2 + \rho \calu
       + \epsilon \frac{m_e}{e^2 n}\frac{|\bs{J}|^2}{2}
       + \frac{|\bs{B}|^2}{2 \mu_0}
     \right) \,,
\end{align*}
is conserved.

Note, if  we were to consider  the governing equations
    (\ref{eq:continuity})--(\ref{eq:adiabatic})
    with the full Maxwell's equations, then the  following energy relation applies:
    \begin{align*}
     0 &= \frac{\p}{\p t}\left(
       \frac{\rho}{2} |\bs{V}|^2 + \rho \calu
       + \epsilon \frac{m_e}{e^2 n}\frac{|\bs{J}|^2}{2}
       + \frac{|\bs{B}|^2}{2 \mu_0} + \frac{\epsilon_0}{2} |\bs{E}|^2
     \right) \nonumber\\
     &~~~~~~
     + \nabla\cdot \left[
      \left( \frac{\rho}{2}|\bs{V}|^2 + p + \rho U
            + \epsilon \frac{m_e}{e^2 n} \frac{|\bs{J}|^2}{2} \right) \bs{V}
       \right.  \nonumber\\
     &~~~~~~~~~~~~~
     \left.
      + \epsilon \frac{m_e}{e^2 n} (\bs{V}\cdot\bs{J})\bs{J}
      - \delta \frac{m_e}{2 e^3 n^2} |\bs{J}|^2 \bs{J}
      + \frac{\bs{E}\times\bs{B}}{\mu_0}
     \right],
    \end{align*}
   and this relation is of the usual conservation form since   $\bs{E}$ is now a dynamical variable.

%%%%%%%%%%%%%%%%%%%%%%%%
%%%%%%%%%%%%%%%%%%%%%%%%
\section{Classification by energy conservation of IMHD}
\label{sec:class-cons}

In this section we  sort  IMHD models into energy conserving  and non-energy conserving classes. 
 
We first consider the compressible IMHD model composed of the 
 pre-Maxwell equations and the following:
\begin{align*}
&\frac{\p \rho}{\p t} + \nabla\cdot(\rho\bs{V})=0, \\
\rho \left( \frac{\p \bs{V}}{\p t}
       + (\bs{V}\cdot\nabla)\bs{V} \right)
  &= - \nabla p + \bs{J}\times\bs{B}
  \nonumber\\
 &\hspace{.35 in}    - \epsilon_{\rm mom}
           \frac{m_e}{e} (\bs{J}\cdot\nabla) \frac{\bs{J}}{en}, \\
\bs{E} + \bs{V}\times\bs{B}
  &= \epsilon_{\rm ti} \frac{m_e}{e^2 n} \frac{\p \bs{J}}{\p t}
     + \epsilon_{\rm ad} \frac{m_e}{e^2 n} (\bs{V}\cdot\nabla)\bs{J}
     \nonumber\\
     &\hspace{-.4 in}+ \epsilon_{\rm cp} \frac{m_e}{e^2 n} \bs{J} (\nabla\cdot\bs{V})
     + \epsilon_{\rm ohm}  \frac{m_e}{e^2 n} (\bs{J}\cdot\nabla)\bs{V}
             \nonumber\\
             &
    \hspace{.25 in} - \delta \frac{m_e}{e^2 n} (\bs{J}\cdot\nabla) \frac{\bs{J}}{en}, \\
 &\frac{\p s}{\p t} + (\bs{V}\cdot\nabla) s=0\,. 
\end{align*}
Recall, the parameters $\delta$ and $\epsilon$ were  artificially inserted into (\ref{eq:continuity})--(\ref{eq:adiabatic}) as  book keeping  parameters,  but now $\epsilon$ has been replaced by  several parameters that track the various effects: $\epsilon_{\rm ti}$ (current time derivative),  $\epsilon_{\rm ad}$ (current advection),
$\epsilon_{\rm cp}$ (compressibility),
$\epsilon_{\rm mom}$ (term in momentum equation that  was $\epsilon$), 
$\epsilon_{\rm ohm}$ (term in Ohm's law partnered with that in the momentum equation).  These parameters are useful for determining how the various terms in the calculation of the energy may  cancel. 

Proceeding, we see the various terms involved in  energy conservation combine  as follows:
\begin{align}
&\frac{\p}{\p t}\left(
       \frac{\rho}{2}  |\bs{V}|^2 + \rho\,  \calu
       + \epsilon_{\rm ti} \frac{m_e}{e^2 n}\frac{|\bs{J}|^2}{2}
       + \frac{|\bs{B}|^2}{2 \mu_0}
     \right) \nonumber\\
&~~~~
   + \nabla\cdot \left[
      \left( \frac{\rho}{2} |\bs{V}|^2 + p + \rho\, \calu
            + \epsilon_{\rm ad} \frac{m_e}{e^2 n} \frac{|\bs{J}|^2}{2}
      \right) \bs{V}
       \right.  \nonumber\\
&~~~~~~~~
     \left.
      + \epsilon_{\rm ohm} \frac{m_e}{e^2 n} (\bs{V}\cdot\bs{J})\bs{J}
      - \delta \frac{m_e}{e^3 n^2} \frac{|\bs{J}|^2}{2} \bs{J}
      + \frac{\bs{E}\times\bs{B}}{\mu_0}
     \right] \nonumber\\
&\hspace{.25 in}
 = (\epsilon_{\rm ti} - \epsilon_{\rm ad})
     \frac{m_e}{e^2 n}
       \frac{|\bs{J}|^2}{2}\frac{\nabla\cdot(n \bs{V})}{n}
  \nonumber\\
  &\hspace{.7in}   + (\epsilon_{\rm ad} - \epsilon_{\rm cp})
      \frac{m_e}{e^2 n} |\bs{J}|^2 (\nabla\cdot\bs{V})
 \label{con1}\\
&\hspace{.95 in}
   + (\epsilon_{\rm ohm} - \epsilon_{\rm mom})
      \frac{m_e}{e} \bs{V} \cdot
       \Big(\bs{J}\cdot\nabla)\frac{\bs{J}}{en} \Big).
      \nonumber
\end{align}
Since our main interest here is with electron inertial models,  we do not consider the case where $\epsilon_{\rm ti}$ vanishes.  Thus, from Eq.~(\ref{con1}) we find that the total energy is conserved only when all the epsilon terms are non-vanishing  or  
$\epsilon_{\rm ti}$, $\epsilon_{\rm ad}$ and $\epsilon_{\rm cp}$ are non-vanishing.  Note,  we have conservation for any value of $\delta$.   Therefore,  we conclude that the epsilon term in the momentum equation is essential for energy conservation in IMHD models.    

Second, we consider the incompressible IMHD model which is governed by the pre-Maxwell equations and by the following equations:
\begin{align*}
\rho_0 \left( \frac{\p \bs{V}}{\p t}
       + (\bs{V}\cdot\nabla)\bs{V} \right)
  &= - \nabla p + \bs{J}\times\bs{B}
  \nonumber\\
  &\hspace{.2 in}
     - \epsilon_{\rm mom}
           \frac{m_e}{e^2 n_0} (\bs{J}\cdot\nabla)\bs{J}, \\
\bs{E} + \bs{V}\times\bs{B}
  &= \epsilon_{\rm ti} \frac{m_e}{e^2 n_0} \frac{\p \bs{J}}{\p t}
     + \epsilon_{\rm ad} \frac{m_e}{e^2 n_0} (\bs{V}\cdot\nabla)\bs{J}
         \nonumber\\
  &~~~~~~
     + \epsilon_{\rm ohm}  \frac{m_e}{e^2 n_0} (\bs{J}\cdot\nabla)\bs{V}
 \nonumber\\
 &\hspace{.7 in}    - \delta \frac{m_e}{e^3 n_0^2} (\bs{J}\cdot\nabla)\bs{J}\,,
\end{align*}
together with $\rho = \rho_0 = {\rm constant}$ or equivalently $n = n_0 = {\rm constant}$. Note that $\epsilon_{\rm cp}$ does not occur in the above equations because of  incompressibility.
For this system,  energy conservation law is as follows: 
\begin{align*}
&\frac{\p}{\p t}\left(
       \frac{\rho_0 }{2}|\bs{V}|^2
       + \epsilon_{\rm ti} \frac{m_e}{e^2 n_0}\frac{|\bs{J}|^2}{2}
       + \frac{|\bs{B}|^2}{2 \mu_0}
     \right) \nonumber\\
&~~~~
   + \nabla\cdot \left[
      \left( \frac{\rho_0}{2} |\bs{V}|^2 + p
            + \epsilon_{\rm ad} \frac{m_e}{e^2 n_0} \frac{|\bs{J}|^2}{2}
      \right) \bs{V}
       \right.  \nonumber\\
&~~~~~~~~
     \left.
      + \epsilon_{\rm ohm} \frac{m_e}{e^2 n_0} (\bs{V}\cdot\bs{J})\bs{J}
      - \delta \frac{m_e}{e^3 n_0^2} \frac{|\bs{J}|^2}{2} \bs{J}
      + \frac{\bs{E}\times\bs{B}}{\mu_0}
     \right] \nonumber\\
&\hspace{.75 in}
 =
   (\epsilon_{\rm ohm} - \epsilon_{\rm mom})
      \frac{m_e}{e^2 n_0} \bs{V} \cdot
      \Big(\bs{J}\cdot\nabla) \bs{J} \Big).
\end{align*}
Thus,  it is revealed that total energy is conserved if  $\epsilon_{\rm ohm} = \epsilon_{\rm mom} = 0$ or 1,
and there are no other conditions on   $\epsilon_{\rm ti}$ and $\epsilon_{\rm ad}$.

\begin{table*}[htbp]
\begin{center}
\begin{tabular}{cccc|c|c|c}
\hline
$\epsilon_{\rm ti}$ & $\epsilon_{\rm ad}$ & $\epsilon_{\rm cp}$
 & $\epsilon_{\rm ohm}$
 & Ohm's law ~ {\scriptsize $\bs{E}+\bs{V}\times\bs{B}=$}
 & $\epsilon_{\rm mom}$ & Conserved?
 \\\hline\hline
 \\
\multicolumn{7}{l}{Compressible plasma} \\\hline
\hline
 {\ }& {\ } & {\ } & {\ }  {\ }& {\ } & {\ } & {\ } \\[-2.5ex]
1 & 1 & 1 & 1
 & {\scriptsize $\dfrac{m_e}{e^2 n}\left(\dfrac{\p \bs{J}}{\p t}
     + \nabla\cdot(\bs{V}\bs{J}+\bs{J}\bs{V})\right)$ }
 & 1 & Yes
 \\[2.0ex]
1 & 1 & 1 &
 & {\scriptsize $\dfrac{m_e}{e^2 n}\left(\dfrac{\p \bs{J}}{\p t}
     + \nabla\cdot(\bs{V}\bs{J})\right)$ }
 &  & Yes
 \\
1 & &  &
 & {\scriptsize $\dfrac{m_e}{e^2 n} \dfrac{\p \bs{J}}{\p t}$}
 &  & {\scriptsize $\dfrac{m_e}{e^2 n} \dfrac{|\bs{J}|^2}{2}
         \dfrac{\nabla\cdot (n\bs{V})}{n} $}
         \\[2.0ex]
1 & 1 &  &
 & {\scriptsize $\dfrac{m_e}{e^2 n}\left(\dfrac{\p \bs{J}}{\p t}
     + (\bs{V}\cdot\nabla) \bs{J} \right)$ }
 &  & {\scriptsize $\dfrac{m_e}{e^2 n} |\bs{J}|^2 (\nabla\cdot\bs{V})$ }
 \\[2.0ex]
1 & 1 & 1 & 1
 & {\scriptsize $\dfrac{m_e}{e^2 n}\left(\dfrac{\p \bs{J}}{\p t}
     + \nabla\cdot(\bs{V}\bs{J}+\bs{J}\bs{V})\right)$ }
 &  & {\scriptsize $\dfrac{m_e}{e} \bs{V} \cdot
       \Big( (\bs{J}\cdot\nabla)\dfrac{\bs{J}}{en} \Big)$ }
 \\[-2.5ex]
  {\ }& {\ } & {\ } & {\ }  {\ }& {\ } & {\ } & {\ } \\
\hline
\\
\multicolumn{7}{l}{Incompressible plasma} \\\hline\hline
 {\ }& {\ } & {\ } & {\ }  {\ }& {\ } & {\ } & {\ } \\[-2.5ex]
1 &  & $-$ &
 & {\scriptsize $\dfrac{m_e}{e^2 n_0} \dfrac{\p \bs{J}}{\p t} $}
 & & Yes
 \\[2.0ex]
1 & 1 & $-$ &
 & {\scriptsize $\dfrac{m_e}{e^2 n_0}\left(\dfrac{\p \bs{J}}{\p t}
     + (\bs{V}\cdot\nabla) \bs{J} \right)$ }
 & & Yes
 \\[2.0ex]
1 & 1 & $-$ & 1
 & {\scriptsize $\dfrac{m_e}{e^2 n_0}\left(\dfrac{\p \bs{J}}{\p t}
     + \nabla\cdot(\bs{V}\bs{J}+\bs{J}\bs{V})\right)$ }
 & 1 & Yes
 \\[2.0ex]
1 & 1 & $-$ & 1
 & {\scriptsize $\dfrac{m_e}{e^2 n_0}\left(\dfrac{\p \bs{J}}{\p t}
     + \nabla\cdot(\bs{V}\bs{J}+\bs{J}\bs{V})\right)$ }
 & & {\scriptsize $\dfrac{m_e}{e^2 n_0} \bs{V} \cdot
       \Big((\bs{J}\cdot\nabla) \bs{J} \Big)$ }
 \\[-2.5ex]
  {\ }& {\ } & {\ } & {\ }  {\ }& {\ } & {\ } & {\ } \\
\hline
\end{tabular}
\caption{
Classification of energy conserving IMHD models. 
The values of the epsilons in the generalized Ohm's law are listed in the first four  columns,
and the generalized Ohm's law is described by  the fifth column.
The epsilon values  in the momentum equation
are listed in the sixth column.
When the total energy is conserved
``Yes'' is written in the last column,
otherwise the deficit terms are written
in the last column.  Note that for  incompressible plasma,
there is no $\epsilon_{\rm cp}$-term in the generalized Ohm's law; consequently, 
 we write $``-"$ in the third column for this case.
}
\label{table:classification_IMHD}
\end{center}
\end{table*}

All of our results on energy conservation for  IMHD models are summarized 
in Table \ref{table:classification_IMHD}.

%%%%%%%%%%%%%%%%%%%%%%%%
%%%%%%%%%%%%%%%%%%%%%%%%
\section{Conclusion}
\label{sec:conclu}

One might argue that the term   $- {m_e}  (\bs{J}\cdot\nabla){\bs{J}}/{(e^2n)}$ of  \eqref{1mom}   may be neglected without consequence,  since it is small.  However, doing so would amount to the introduction of nonphysical dissipation.  Since the whole point of reconnection studies is that a small physical dissipation can have important consequences,  one should view any reconnection calculation with this nonphysical dissipation with caution.   Note, however, in some geometries this term  may vanish.

Equations (3.7.1)--(3.7.4) of Ref.~\cite{KT}  do  not conserve  energy, whether or not Maxwell's displacement current is retained.  In this reference and elsewhere it is  described how the  neglected term 
%\[
% (\bs{J}\cdot\nabla)({\bs{J}}/{n})= \nabla\cdot (\bs{J} \otimes{\bs{J}}/{n})
%\]
in the momentum equation can be recognized by reverting to a two-species kinetic theory, where the pressure tensors for each species are given by
\bq
 \underline{\underline{P}}^{\alpha}= m_{\al}\int\!d^3v\, f_{\al}\, (\bfv- \bfv_{\al})\otimes(\bfv- \bfv_{\al})
 \eq
 with $\al\in\{e,i\}$, $f_{\al}$ being  the phase space density of species $\al$, and the fluid velocities  given as usual by 
 \bq
 \bfv_{\al}=  \frac{\int\!d^3v\, f_{\al}\, \bfv}{\int\!d^3v\, f_{\al}}\,.
 \eq
 If we follow L\"ust's example for scalar pressure, then the total pressure is the sum of the partial pressures, i.e.,   
 \bq
  \mathbb{P}=    \mathbb{P}^{i} +    \mathbb{P}^{e}\,.
  \label{P}
  \eq
 in accordance with Dalton's law. However, it is sometimes suggested that one use pressures defined in terms of the center of mass velocity according to
 \bq
  \mathbb{P}^{\alpha}_{\rm cm}= m_{\al}\int\!d^3v\, f_{\al}\, (\bfv- \bfV)\otimes(\bfv- \bfV)
 \eq
 and define the total pressure by  
 $ \mathbb{P}_{\rm cm}=   \mathbb{P}_{\rm cm}^{i} 
  +    \mathbb{P}_{\rm cm}^{e}$. 
Upon inserting
 \bq
 \bfv_i=\bfV +\frac{m_e}{m} \frac{\bfj}{e n}\quad\mathrm{and} \quad \bfv_e=\bfV - \frac{m_i}{m} \frac{\bfj}{e n}
 \eq
 in (\ref{P}), an easy calculation gives
 \bq
   \mathbb{P}=    \mathbb{P}_{\rm cm} -\frac{m_em_i}{me^2n}\, \bfj \otimes\bfj
  \approx   \mathbb{P}_{\rm cm} -\frac{m_e}{e^2n}\, \bfj \otimes\bfj
 \eq
 Thus, one could replace the first and last terms on the right hand side of (\ref{1mom}) by 
 $-\nabla\cdot \mathbb{P}_{\rm cm}$ and obtain a tidy equation.  However, if one further makes conventional thermodynamic closure assumptions on $\mathbb{P}_{\rm cm}$, e.g., that it is isotropic and  either  barotropic or adiabatic, then one would be in essence saying that the current is dependent on density, which is unphysical.  This unphysical nature is manifest in the resulting violation of energy conservation when this procedure is employed. 
 
In this paper we have used physical reasoning and direct calculation to obtain   conserved energy densities.  However,  energy should emerge from time translation symmetry by means of Noether's theorem.  That this is indeed the case will be reported in future work  \cite{klmww} by deriving the action for extended MHD and then using the Galilean group to construct the usual conservation laws.   With this formalism one also obtains the noncanonical Poisson brackets for this model akin to that of  \cite{MG80} (see \cite{rmp98} for review).   This leads to  the Casimir invariants and opens up the possibility of applying Hamiltonian techniques for stability such as in \cite{amp0,amp1,amp2,mma14}.

Finally, we point out that our starting point was two-fluid theory and gyroviscous effects due to strong  magnetic fields have not been incorporated  \cite{HH71,RT62,mma14}.   This will also be the subject of a future publication.

\bigskip

\section*{Acknowledgements}

We would like to acknowledge  support and hospitality of the 2011 Geophysical Fluid Dynamics Program held at the  Woods Hole Oceanographic Institution,  where the bulk of this research was undertaken.    K.K.\  was  supported by a Grant in-Aid from  the Global COE Program  ``Foundation of International Center for Planetary Science''
from the Ministry of Education, Culture, Sports, Science and Technology (MEXT) of Japan.  P.J.M.\  would like to acknowledge useful conversations with Francesco Pegoraro;  he was supported by U.S. Dept.\ of Energy Contract \# DE-FG05-80ET-53088.

\bibliographystyle{pf}

\end{document}